\documentclass[useAMS,usenatbib]{mn2e}

\usepackage{graphicx} 

\title[Jet and disc outflows in 3C~111]{Comparison of ejection events in the jet and accretion disc outflows in 3C~111}
\author[F. Tombesi et al.]{F. Tombesi$^{1,2}$\thanks{E-mail: ftombesi@astro.umd.edu}, R. M. Sambruna$^{3}$, A. P. Marscher$^{4}$, S. G. Jorstad$^{4,5}$, C. S. Reynolds$^2$\newauthor and A. Markowitz$^{6,7,8}$\\
$^{1}$X-ray Astrophysics Laboratory and CRESST, NASA/Goddard Space Flight Center, Greenbelt, MD 20771, USA\\
$^{2}$Department of Astronomy, University of Maryland, College Park, MD 20742, USA\\
$^{3}$Department of Physics and Astronomy, MS 3F3, 4400 University Drive, George Mason University, Fairfax, VA 22030, USA\\
$^{4}$Institute for Astrophysical Research, Boston University, 725 Commonwealth Ave., Boston, MA 02215, USA\\
$^{5}$Astronomical Institute, St. Petersburg State University, Universitetskij Pr. 28, Petrodvorets, 198504 St. Petersburg, Russia\\
$^{6}$Center for Astrophysics and Space Sciences, University of California, San Diego, Mail Code 0424, La Jolla, CA 92093-0424, USA\\
$^{7}$Dr. Karl Remeis-Sternwarte and Erlangen Centre for Astroparticle Physics, Frederic-Alexander
 Universit\"{a}t Erlangen-N\"{u}rnberg, 7\\ Sternwartstra{\ss}e, 96049 Bamberg, Germany\\
$^{8}$Alexander von Humboldt Fellow}
\begin{document}

\date{Accepted ???. Received ???; in original form ???}


\maketitle

\label{firstpage}

\begin{abstract}

We present a comparison of the parameters of accretion disc outflows and the jet of the broad-line radio galaxy 3C~111 on sub-pc scales. We make use of published X-ray observations of ultra-fast outflows (UFOs) and new 43~GHz VLBA images to track the jet knots ejection. We find that the superluminal jet coexists with the mildly relativistic outflows on sub-pc scales, possibly indicating a transverse stratification of a global flow. The two are roughly in pressure equilibrium, with the UFOs potentially providing additional support for the initial jet collimation. The UFOs are much more massive than the jet, but their kinetic power is probably about an order of magnitude lower, at least for the observations considered here. However, their momentum flux is equivalent and both of them are powerful enough to exert a concurrent feedback impact on the surrounding environment. A link between these components is naturally predicted in the context of MHD models for jet/outflow formation. However, given the high radiation throughput of AGNs, radiation pressure should also be taken into account.
From the comparison with the long-term 2--10~keV RXTE light curve we find that the UFOs are preferentially detected during periods of increasing flux. We also find the possibility to place the UFOs  within the known X-ray dips-jet ejection cycles, which has been shown to be a strong proof of the disc-jet connection, in analogue with stellar-mass black holes. However, given the limited number of observations presently available, these relations are only tentative and additional spectral monitoring is needed to test them conclusively.

\end{abstract}


\section{Introduction}

One of the most enduring open questions surrounding active galactic nuclei (AGNs) concerns the relation between accretion and ejection processes, i.e., what is the connection between the black hole, the accretion disc and the formation of outflows and jets? Then, a related question is, what is the feedback impact of AGN outflows/jets on the host galaxy and surrounding environment? 
New insights on the characteristics and importance of winds/outflows in radio-quiet AGNs have been recently obtained thanks to deep \emph{XMM-Newton}, \emph{Chandra} and \emph{Suzaku} observations. In particular, the detection of blue-shifted Fe~XXV-XXVI absorption lines in the X-ray spectra of several sources demonstrated the presence of massive, highly ionized and mildly/nearly relativistic accretion disk outflows (e.g., Chartas et al. 2002, 2003; Pounds et al.~2003; Dadina et al.~2005; Markowitz et al.~2006; Braito et al.~2007; Cappi et al.~2009; Reeves et al.~2009; Chartas et al.~2009a; Giustini et al.~2011; Dauser et al.~2011; Gofford et al.~2011; Lobban et al.~2011). Moreover, a systematic spectral analysis and photo-ionization modelling performed by Tombesi et al.~(2010a; 2011a; 2012) on a sample of 42 Seyfert galaxies observed with \emph{XMM-Newton} demonstrated that these ultra-fast outflows (UFOs) are rather common phenomena, being present in more than 40\% of the sources, and confirmed the claims that these UFOs are indeed powerful enough to potentially play a significant role in the AGN cosmological feedback.

\begin{table*}
\centering
\begin{minipage}{150mm}
\caption{\emph{Suzaku} and \emph{XMM-Newton} observations of 3C~111 and parameters of the detected UFOs.}
\begin{tabular}{@{\hspace{0.2cm}}l@{\hspace{0.2cm}}c@{\hspace{0.2cm}}c@{\hspace{0.2cm}}c@{\hspace{0.2cm}}cc@{\hspace{0.2cm}}c@{\hspace{0.2cm}}c@{\hspace{0.2cm}}c@{\hspace{0.2cm}}c@{\hspace{0.2cm}}c@{\hspace{0.2cm}}}
\hline
 & Sat & T$_{\mathrm{obs}}$ & UFO & log$L_{\mathrm{ion}}$ & log$N_\mathrm{H}$ & log$\xi$ & $v_{\mathrm{out}}$ & $r$ & $\dot{M}_{\mathrm{out}}$ & log$\dot{E}_\mathrm{K}$ \\
    &     &    &  & (erg~s$^{-1}$) & (cm$^{-2}$) & (erg~s$^{-1}$~cm) & (c) & (pc) & ($M_{\odot}$~yr$^{-1}$) & (erg~s$^{-1}$) \\
\hline
1$^*$ & S & 2008.65 & u7 & 44.4 & $>23.0$ & $5.0\pm0.3$ & $0.041\pm0.003$ & 0.003-0.02 & 0.1-0.6 & 42.8-43.5\\
2 & X & 2009.13  & & 44.7 & \dots & \dots & \dots & \dots & \dots & \dots \\
3 & S & 2010.67 &  & 44.7 & \dots & \dots & \dots & \dots & \dots & \dots \\
4$^*$ & S & 2010.69 & u10 & 44.9 & $22.9\pm0.2$ & $4.3\pm0.1$ & $0.106\pm0.006$ & 0.001-0.006 & 0.1-0.8 & 43.4-44.3\\
5 & S & 2010.71 & & 44.9 & \dots & \dots & \dots & \dots & \dots & \dots \\
\hline
\end{tabular}
{\em Note.} Columns: observation number; satellite, X for \emph{XMM-Newton} and S for \emph{Suzaku}; observation start date; UFO identifier; absorption corrected luminosity between 1--1000~Ryd (1~Ryd=13.6~eV); column density; ionization parameter; observed velocity; location; mass outflow rate; kinetic power. $^*$ Observations with detected UFOs.
\end{minipage}
\end{table*}

In radio-loud AGNs, relativistic jets are routinely observed at radio, optical and X-rays. However, the presence of disc outflows in these objects has recently emerged thanks to X-ray spectroscopy. For instance, Tombesi et al.~(2010b) reported the discovery of highly ionized and massive gas outflowing with mildly relativistic velocities $\sim$0.1c, consistent with UFOs, in 3/5 Broad-Line Radio Galaxies (BLRGs) observed with \emph{Suzaku}, namely 3C~111, 3C~120 and 3C~390.3. The UFO in 3C~111 was detected in a long observation performed in August 2008 (see Table~1) and a follow-up study was then performed in September 2010 to study its variability through three \emph{Suzaku} observations spaced by $\sim$7~days (Tombesi et al.~2011b). A systematic 4--10~keV spectral analysis revealed the presence of an ionized Fe K emission line in the first observation, indicative of reflection/emission from the accretion disc, and blue-shifted Fe K absorption in the second one, when the flux was 30\% higher, indicating the presence of a highly ionized and massive outflow with velocity $\sim$0.1c (see Table~1). The location of the material was constrained at $\la$0.006~pc ($\la$500$r_\mathrm{s}$, $r_\mathrm{s}$$=$$2GM_{\mathrm{BH}}/c^2$) from the black hole through the $\sim$7~days variability time-scale. This provided the first direct evidence for an accretion disc-outflow connection in an AGN and is consistent with a picture in which a disruption/over-ionization of the inner disc is followed by the ejection of an outflow from $\sim$100$r_\mathrm{s}$. Then, this is possibly accelerated through radiation and/or magnetic forces to the observed velocity of $\sim$0.1c.

Chatterjee et al.~(2011; hereafter Ch11) recently reported the results of an extensive multi-frequency monitoring campaign on 3C 111 at X-ray (2--10 keV), optical (R band), and radio (14.5, 37 and 230 GHz) wave bands, as well as multi-epoch imaging with the Very Long Baseline Array (VLBA) at 43 GHz, between 2004 and 2010.
They find that major X-ray dips are systematically followed by an increase of the radio core flux and the appearance of new jet knots in the VLBA images after $\sim$2--3 months. New knots are ejected $\sim$1--2 times per year with typical apparent speeds of $\sim$3--5c. This shows the existence of a connection between the radiative state near the black hole, where the X-rays are produced, and events in the jet, providing a solid proof of the disc-jet connection in this radio-loud AGN. 
These complex cycles, whereby some instability appears to disrupt the inner regions of the accretion disc and then triggers powerful mass/energy ejections provide an observational clue to the origin of radio jets. This behaviour has parallels in Galactic microquasars (e.g., Neilsen \& Lee 2009; Fender et al.~2009) and equivalent results have been obtained also for another BLRG, 3C~120 (Marscher et al.~2002; Chatterjee et al.~2009).

The BLRG 3C~111 ($z$$=$0.0485) is one of the best targets for these studies. In the radio it is an FR~II with a blazar-like behaviour. The jet lies at $\theta$$=$$18.1^{\circ}$$\pm$$5.0^{\circ}$ to our line of sight (Jorstad et al.~2005), allowing us to simultaneously probe the inner accretion disc through X-rays. 3C~111 is X-ray bright (2--10~keV flux of $\sim$$2-8\times 10^{-11}$~erg~s$^{-1}$~cm$^{-2}$) and shows Seyfert-like properties. It is also one of the two BLRGs, the other being 3C~120, recently detected in $\gamma$-rays with \emph{FERMI} (Kataoka et al.~2011; Grandi et al.~2012).
For the central black hole of 3C~111 we consider a mass\footnote{We note that Marchesini et al.~(2004) estimated a larger black hole mass of log$M_{\mathrm{BH}}$$\sim$9.5~$M_{\odot}$ assuming the bulge luminosity relation. The discrepancy is probably mainly due to the different extinction adopted and we consider the Ch11 one to be more reliable.} of log$M_{\mathrm{BH}}$$=$$8.1\pm0.5$~$M_{\odot}$, taking into account the maximum and minimum values derived by Ch11 using H$\alpha$ measurements. The Eddington luminosity is therefore $L_{\mathrm{Edd}}$$=$$1.3\times 10^{38} (M_{\mathrm{BH}}/M_{\odot})$$\simeq$$2\times 10^{46}$~erg~s$^{-1}$. 

In this work we will focus on the comparison between ejection events in the radio jet and those from the accretion disc of 3C~111, represented by the UFOs, and the search for a possible link between these two.
The paper is structured as follows. In \S~2 we estimate the parameters of the UFOs in 3C~111 using the published data. In \S~3 we extend the work of Ch11 and estimate the parameters of the inner radio jet from the VLBA images. In \S~4 we compare the characteristics of the UFOs and the jet and discuss the possibility to place also the UFOs in the context of the source variability and the known jet ejection cycles, with conclusions following in \S~5. Throughout this paper we adopt a Hubble constant of $H_0$$=$$70$~km~s$^{-1}$~Mpc$^{-1}$ (Spergel et al.~2003).

\section[]{Observations of UFOs}

\subsection[]{Parameters from X-ray spectroscopy}

Table~1 reports the parameters of the published \emph{Suzaku} and \emph{XMM-Newton} observations of 3C~111 in which a search for UFOs has been performed. The first observation refers to Tombesi et al.~(2010b), the second to Ballo et al.~(2011) and the last three to Tombesi et al.~(2011b). The column density, ionization and outflow velocity are reported.   
We estimate the lower/upper limits of the location, mass outflow rate and kinetic power of the UFOs following the approach of Tombesi et al.~(2012). 

An estimate of the minimum distance can be derived from the radius at which the observed velocity corresponds to the escape velocity, $r_{\mathrm{min}} = 2 G M_{\mathrm{BH}}/ v_{\mathrm{out}}^{2}$. However, we note that when deriving this quantity we do not take into account the possible additional acceleration of the flow, but assume that it is ejected at the observed velocity.
Instead, in order to derive a firm estimate of the distance from the definition of the ionization parameter $\xi=L_{\mathrm{ion}}/nr^2$ (Tarter et al.~1969) we would need an estimate of the density of the material $n$, which can not be obtained with the present data. However, the observed short time-scale variability of the UFOs (e.g., Braito et al.~2007; Cappi et al.~2009; Tombesi et al.~2010a, 2011b; Giustini et al.~2011) suggests that these absorbers are compact and that their thickness is less than the distance to the source, $\Delta r/r$$<$1. Therefore, we can derive a lower limit of the density of the material as $n$$=$$N_{\mathrm{H}}/ \Delta r$$>$$N_{\mathrm{H}}/r$, where $N_{\mathrm{H}}$ is the line-of-sight column density. Then, substituting this in the expression for the ionization parameter we can estimate an upper limit on the distance of the absorber from the source as $r_{\mathrm{max}} = L_{\mathrm{ion}}/\xi N_\mathrm{H}$. The material can not be farther away than this given the observed ionization parameter.
We note that this expression contains the implicit assumption that the ionizing source is seen as a point source by the absorber. The validity of this supposition is supported from the fact that the X-ray emitting region in AGNs is constrained by X-ray variability and micro-lensing observations to be within a few $r_\mathrm{s}$ from the black hole (e.g., Chartas et al. 2009b), instead the UFOs considered here are always at distances $\ga$100$r_{\mathrm{s}}$ (see text below and Table~1).

  \begin{figure*}
  \centering
   \includegraphics[width=18cm,height=13cm,angle=0]{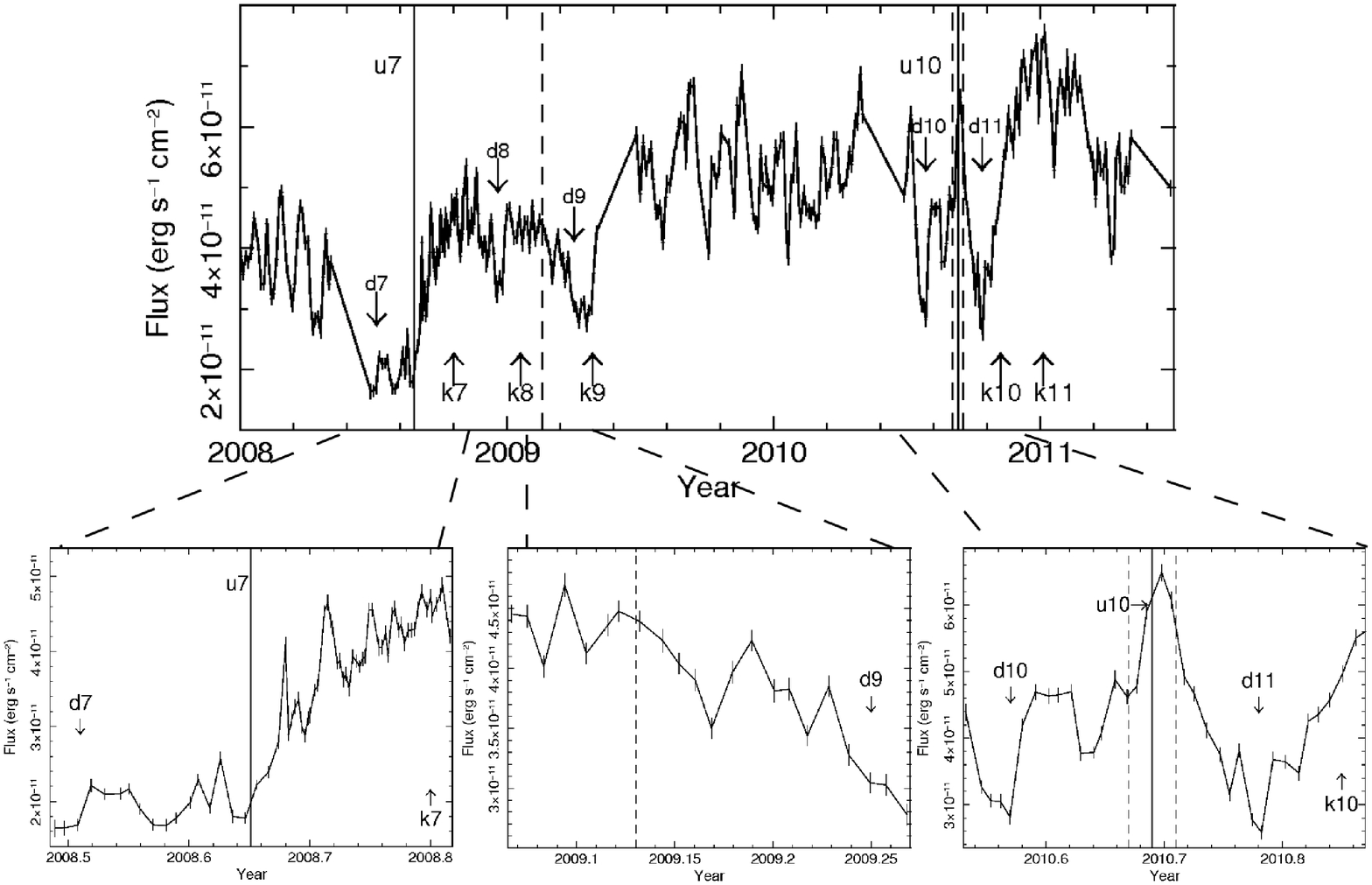}
   \caption{Long-term 2.4--10~keV flux RXTE light curve of 3C~111 between 2008 and mid 2011. The vertical solid/dotted lines refer to the detection/non-detection of UFOs in the \emph{Suzaku} and \emph{XMM-Newton} spectra. The detections of UFOs are marked with ``u''. The dates relative to the X-ray dips and the appearance of new jet knots in the VLBA images are marked with ``d'' and ``k'', respectively.}
    \end{figure*}

We use the expression for the mass outflow rate derived by Krongold et al.~(2007), which is more appropriate for a biconical wind-like geometry: $\dot{M}_{\mathrm{out}}\simeq1.2 \pi m_\mathrm{p} N_\mathrm{H} v_{\mathrm{out}} r$. This formula has also the important advantage of already implicitly taking into account the covering and filling factors. This is due to the fact that it considers only the net flow of mass, directly allowing for clumping in the flow. 
However, we note that considering a clumpiness factor of $\sim$$\Delta r/r$ we obtain equivalent results using the usual spherical approximation (Tombesi et al.~2010b, 2011b) and a covering fraction of $\sim$0.3, consistent with observations (Tombesi et al.~2010a,b). The kinetic power can consequently be derived as $\dot{E}_\mathrm{K} = \frac{1}{2} \dot{M}_{\mathrm{out}} v_{\mathrm{out}}^2$.

Substituting the relative parameters, the UFO observed in August 2008, u7 in Table~1, is constrained at a distance\footnote{Assuming the higher black hole mass estimate of Marchesini et al.~(2004), as in Tombesi et al.~(2010a), the observed velocity of u7 would be lower than the escape velocity at the estimated location. However, using the refined lower black hole mass estimate of Ch11 (see \S1) and considering also the possibility of additional acceleration, we are confident enough that also u7 escapes the system.} of $\simeq$0.003--0.02~pc ($\simeq$200--1000$r_\mathrm{s}$) from the central black hole, with a mass outflow rate of $\dot{M}_{\mathrm{out}}$$\simeq$0.1--0.6~$M_{\odot}$~yr$^{-1}$ and kinetic power of $\dot{E}_\mathrm{K}$$\simeq$$6\times 10^{42}$--$3\times 10^{43}$~erg~s$^{-1}$. The UFO relative to the September 2010 observation, u10 in Table~1, is located at $\simeq$0.001--0.15~pc ($\simeq$80--12,000$r_\mathrm{s}$), it has an outflow rate of $\dot{M}_{\mathrm{out}}$$\simeq$0.1--10~$M_{\odot}$~yr$^{-1}$ and a kinetic power of $\dot{E}_\mathrm{K}$$\simeq$$2\times 10^{43}$--$4\times 10^{45}$~erg~s$^{-1}$. However, considering the variability on $\sim$7~days time-scales (Tombesi et al.~2011b), we can further constrain its distance to $\la$0.006~pc ($\la$500$r_\mathrm{s}$). Consequently, the upper limits on the mass outflow rate and kinetic power for u10 reduce to $\la$0.8~$M_{\odot}$~yr$^{-1}$ and $\la$$2\times 10^{44}$~erg~s$^{-1}$, respectively. Even though u10 is faster than u7, we note that many of their other characteristics ($N_\mathrm{H}$, $r$, $\dot{M}_{\mathrm{out}}$ and $\dot{E}_\mathrm{K}$) are consistent with each other.
 
From the relation $L_{\mathrm{bol}}$$\simeq$$10 L_{\mathrm{ion}}$~erg~s$^{-1}$ (e.g., McKernan et al.~2007) the bolometric luminosity is $L_{\mathrm{bol}}$$\simeq$$10^{45}$~erg~s$^{-1}$, which corresponds to an Eddington ratio of $L_{\mathrm{bol}}/L_{\mathrm{Edd}}$$\simeq$0.1. Considering a radiative efficiency of $\eta$$\simeq$0.1 (Davis \& Laor 2011), the accretion rate is $\dot{M}_{\mathrm{acc}}$$=$$L_{\mathrm{bol}}/\eta c^2$$\simeq$0.5~$M_{\odot}$~yr$^{-1}$, which is comparable to the outflow rate derived for the UFOs.

\subsection[]{Variability and the X-ray light curve}

In Fig.~1 we show the 2.4--10~keV flux light curve of 3C~111 observed with the \emph{Rossi X-ray Timing Explorer} (RXTE) from the beginning of 2008 to mid 2011. The typical exposure time is 1--2~ks and the sampling of the observations was 2--3 times per week. We adopted the same data reduction procedures as explained in Ch11. We observe five major dips in the light curve and, adopting the same nomenclature of Ch11, we identify them in Table~2 with ``d'' and the relative number. 

We can use this light curve to check for possible relations between the source flux variability and that of the UFOs. There are three possible causes for the observed variability of UFOs. 
First, it could be an intermittent on/off variability, i.e. the UFOs are not a continuous phenomenon and they are ejected only at certain times.  In this case, the lack of detection is due to the absence of a UFO at the time of the observation.
Second, even if a UFO is present during the time of the observation, there could be some additional absorber variability due to inhomogeneities in the structure and density of the flow (e.g., turbulence, clumpiness) and transverse motion. This effect is expected to occur on much shorter time-scales than the first one, on intervals of the order of a few hours (e.g., Braito et al.~2007; Giustini et al.~2011). This is also expected to be chaotic and not correlated with a state of the source. Third, the absence of a UFO could be masked by an insufficient signal-to-noise (S/N) of the data. 

The UFOs have been clearly detected in 2/5 observations, thus their frequency of detection is $\sim$40\%. In particular, the F-test and Monte Carlo probabilities for the absorption lines detected in both the first (Tombesi et al.~2010b) and fourth observations (Tombesi et al.~2011b) in Table~1 are $>$99\%. In addition, as an alternative test on the significance of the lines (e.g., Vaughan \& Uttley 2008), we note that the ratio between the equivalent width and the relative 1$\sigma$ errors\footnote{We note that the equivalent width errors of the absorption lines reported in Table~3 of Tombesi et al.~(2010b) for 3C~111 are at the 90\% level, instead to the 1$\sigma$ level indicated in the table notes.} is $\ga$4 for all the cases.
As stated by Ballo et al.~(2011), the non-detection of a UFO in the second observation in Table~1 is not due to a low S/N. The same conclusion is derived also for the third and fifth observations by Tombesi et al.~(2011b). In fact, the 2--10~keV S/N in these observations was $\simeq$50, 190, 105, 110 and 105 for the first, second, third, fourth and fifth observation, respectively. Therefore, this indicates that a UFO was not present along the line-of-sight during these observations or it was so weak that it could not be detected even in high S/N spectra. 

We then performed a test in order to check for a possible relation between the source X-ray flux from the RXTE light curve shown in Fig.~1 and the detections/non-detections of UFOs (marked with solid/dotted vertical lines, respectively). We can see that there seems to be no dependence on the absolute flux of the source, the UFOs being detected/non-detected both in high/low flux states. However, there could be a relation with the flux variability trends.
In order to distinguish between periods of rising or steady/decreasing flux we consider the difference between the fluxes $\sim$3~days after and before the observations. This is equivalent to the minimum time interval of the RXTE light curve and also allows to oversample by a factor of $\sim$2 the typical variability time-scale of the UFO in 3C~111 of about $\sim$7~days (Tombesi et al.~2010b, 2011b). If the difference is positive it indicates a rising flux, instead if null or negative it indicates a steady/decreasing flux. We find that the first and fourth observations in Table~1, the ones with detected UFOs, happened during periods of increasing flux\footnote{We note that the observation of the UFO u7 occurred at the beginning of a period of rising flux, just after the major X-ray dip d7. If the rising period is related with the acceleration of the outflow, this might explain why the velocity of u7 is much lower than u10, which instead was detected close to a maximum in flux.}. Instead, the non-detections in the second and fifth observations occurred during intervals of decreasing flux. Following this criterion, the non-detection in the third observation occurred in an interval of steady/decreasing flux too. However, we note that this latter case is less stringent because it happened very close to a sudden spike in flux and we adopt a conservative approach not considering it in the following discussion. 

From Fig.~1, we derive that overall the UFOs seem to be preferentially detected during intervals of increasing flux. In order to estimate the statistical confidence of the possible relation between the UFOs and the periods of rising flux we tested the null hypothesis that UFOs are not detected during phases of ascending flux but only in steady or decreasing intervals. This hypothesis is satisfied in none of the four cases described before, yielding a probability of $<$1/4. Therefore, conservatively, we can say that the statistical probability of the claim that UFOs are preferentially observed during phases of rising flux is $P$$=$$1-(1/4)$$\ga$75\%. Given the limited number of observations available, we stress that the statistical significance of this relation is only marginal and it should be regarded only as an indication. However, we note that a similar behavior was observed also in other sources showing UFOs (e.g., Braito et al.~2007; Giustini et al.~2011).

\begin{table}
\caption{Times of X-ray dips, observations of UFOs and appearance of radio knots.}
\begin{tabular}{@{\hspace{0.2cm}}l@{\hspace{0.2cm}}c@{\hspace{0.2cm}}l@{\hspace{0.2cm}}c@{\hspace{0.2cm}}l@{\hspace{0.2cm}}c@{\hspace{0.2cm}}c@{\hspace{0.2cm}}}
\hline
Dip & T$_{\mathrm{Xmin}}$ & UFO & T$_{\mathrm{ufo}}$ & Knot & T$_{\mathrm{knot}}$ & $\beta_{\mathrm{app}}$ \\
\hline
d7 & 2008.51 & u7 & 2008.65 & k7 & $2008.83\pm0.07$ & $4.54\pm0.38$\\
d8 & 2008.98 & \dots & \dots & k8 & $2009.07\pm0.08$ & $4.07\pm0.43$\\
d9 & 2009.26 & \dots & \dots & k9 & $2009.29\pm0.04$ & $4.33\pm0.66$\\
d10 & 2010.57 & u10 & 2010.69 & k10 & $2010.85\pm0.02$ & $5.66\pm0.09$\\
d11 & 2010.78 & \dots & \dots & k11 & $2011.01\pm0.07$ & $5.22\pm0.35$\\
\hline
\end{tabular}
{\em Note.} $\beta_{\mathrm{app}}$ is the apparent speed of the radio knots in units of c.
\end{table}

\section[]{Radio observations of the jet on sub-pc scales}

3C~111 is actively monitored with the VLBA at 43~GHz at roughly monthly intervals by the blazar group at the Boston University. Here we present a temporal extension of the VLBA analysis of Ch11 (see their Fig.~6) from 2008 up to mid 2011. The sequence of VLBA images shown in Fig.~2 provides a dynamic view of the inner jet between November 2010 and September 2011 at an angular resolution $\sim$0.1 milliarcseconds (mas), corresponding to $\sim$0.094~pc. The VLBA data have been processed in the same manner as described in Ch11. 
We can clearly observe the presence of two new radio jet knots, each characterized by its flux density, FWHM diameter and position relative to the core. Times of ``ejection'' are defined as the extrapolated time of coincidence of a moving knot with the position of the 43~GHz core.
We use the position vs. time data to determine the projected direction on the sky of the inner jet, as well as the apparent speeds and ejection times of new superluminal knots. Continuing with the nomenclature adopted by Ch11, we have knot k10 appearing from the 43~GHz core at $2010.85\pm0.02$ and k11 at $2011.01\pm0.07$, respectively. Both of them have apparent superluminal velocities of $5.66\pm0.09$c and $5.22\pm0.35$c, respectively. The proper motion of these knots can be directly followed in Fig.~2 for almost one year. 

The dates relative to the X-ray dips and jet knots appearance between 2008 and 2011 are marked in the RXTE light curve in Fig.~1 by arrows. We see that, in line with the reported correlation between X-ray dips and jet ejections (Ch11), new radio jet knots systematically appear a few months after major X-ray dips. 
This is valid also for the two new detected ones, k10 and k11, which appear about $\sim$3 months after the relative X-ray dips d10 and d11. We are confident that the dips d10 and d11 are indeed related to the knots k10 and k11 for several reasons: there is no significant detection of a new knot ejection in the radio images between 2010.60 and 2010.80, the time interval between d10 and d11 is equivalent to that between k10 and k11 and also both knots appear in the radio images with an equivalent delay of about 3 months after the relative dips. In general, the delay is distributed between 0.03~yr and 0.34~yr, with a mean value of $0.15\pm0.08$~yr (Ch11).   
As already discussed by Ch11, considering the apparent speeds of $\sim$4--5c, an average delay from the X-ray dips of $\sim$0.15~yr and an inclination of $\sim$18$^{\circ}$, we can derive that the typical distance traveled by the jet knots before appearing out from the 43~GHz core is $d$$\sim$0.6pc.

Considering the jet knots k7 and k10, their actual bulk velocity ($v$$=$$\beta c$) can be estimated from the apparent velocity ($v_{\mathrm{app}}$$=$$\beta_{\mathrm{app}} c $) adopting an inclination to the line of sight of $\theta$$\sim$18$^{\circ}$ (Jorstad et al.~2005), $\beta_{\mathrm{app}} = \beta \mathrm{sin}\theta (1 - \beta\mathrm{cos}\theta)^{-1}$. We obtain $v_{\mathrm{k7}}$$\simeq$0.982c and $v_{\mathrm{k10}}$$\simeq$$0.995$c and the relative Lorentz factors ($\Gamma$$=$$1/\sqrt{1- \beta^2}$) are $\Gamma_{\mathrm{k7}}$$\sim$5.3 and $\Gamma_{\mathrm{k10}}$$\simeq$10 for k7 and k10, respectively. The knot k10 is faster than k7 and their parameters are reported in Table~2. 

Assuming equipartition and using the formula (A3) in Jorstad \& Marscher (2004) we derive an estimate of the magnetic field of $B$$\simeq$0.1~Gauss, consistent with the typical values at $\sim$sub-pc scales (e.g., O'Sullivan \& Gabuzda 2009). Then, it is possible to roughly quantify the jet kinetic power as $\dot{E}_{\mathrm{K,j}}$$\simeq$$2(B^2/8\pi)(\pi d^2)c$$\sim$$3\times 10^{44}$~erg~s$^{-1}$, where $d$ is the previously estimated distance of the knots from the black hole. However, if we include also the possible additional term due to the rest-mass energy of the protons, the kinetic power can reach values up to $\sim$$10^{45}$~erg~s$^{-1}$. These estimates are consistent with the typical jet power of radio galaxies estimated from the associated radio lobes of $\dot{E}_{\mathrm{K, j}}$$\sim$$10^{44}$--$10^{45}$~erg~s$^{-1}$ (e.g., Rawlings \& Saunders 1991). 
Subsequently, from the relation $\dot{E}_{\mathrm{K,j}}$$\simeq$$(1/2) \dot{M}_{\mathrm{out,j}} c^2 (\Gamma - 1)$ we can also calculate the mass flux rate that is funneled into the jet. Considering the average $\Gamma$$\sim$7 of the jet knots in Table~2, we obtain a mass outflow rate of $\dot{M}_{\mathrm{out, j}}$$\simeq$0.0005--0.005~$M_{\odot}$~yr$^{-1}$.

  \begin{figure}
  \centering
   \includegraphics[width=7.5cm,height=22cm,angle=0]{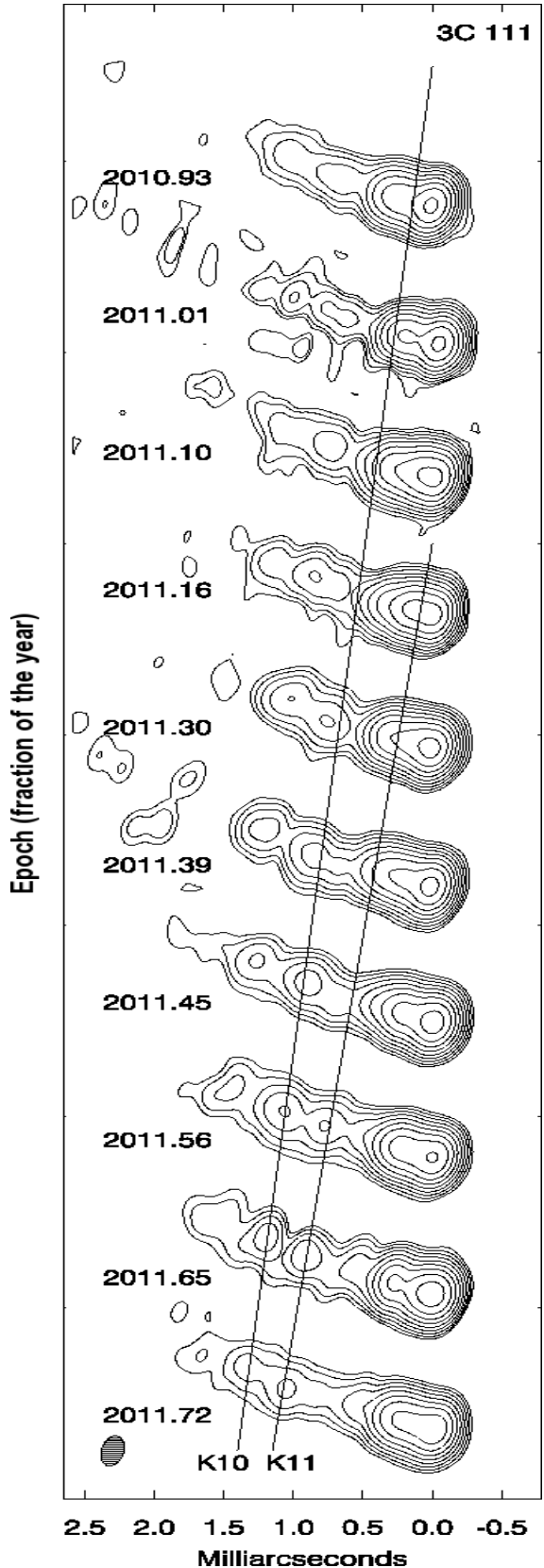}
   \caption{Sequence of VLBA images at 43~GHz during 2010--2011. The global peak of the map is 1.13 Jy/beam with the beam size of 0.32$\times$0.16 mas$^2$ at PA=-10$^\circ$, the contours levels start at 0.25\% of the peak and increase by a factor of 2. The lines denote the proper motion of the radio jet knots k10 and k11.}
    \end{figure}

\section[]{Discussion}

In this paper we focus on a comparison between the parameters of the jet and accretion disc outflows, also referred as UFOs, observed in 3C~111. This is the first time that such a study is performed for an AGN.
The typical kinetic power of the UFOs reported in Table~1 is $\dot{E}_\mathrm{K}$$\sim$$10^{43}$--$10^{44}$~erg~s$^{-1}$. This is about one order of magnitude lower than the kinetic power of the jet. 
However, if we take into account the uncertainty in the determination of the jet kinetic power and the possibility that the jet knots experienced an additional acceleration with respect to the UFOs, it is plausible that their values are actually comparable. 
In fact, the UFOs were detected at distances of only $\sim$0.001--0.02~pc from the black hole, well within the $\sim$0.6~pc scale of the 43GHz core, after which new jet knots are observable in the VLBA images. 

The mass outflow rate of the UFOs of $\dot{M}_{\mathrm{out}}$$\sim$0.1--1~$M_{\odot}$~yr$^{-1}$ is much higher than that estimated for the jet $\dot{M}_{\mathrm{out, j}}$$\sim$0.0005--0.005~$M_{\odot}$~yr$^{-1}$. Moreover, even if their kinetic power is different, their momentum flux is actually equivalent. This is due to the linear dependence of this quantity on the outflow velocity. 
The ratio of the kinetic power with respect to the bolometric luminosity corresponds to $\sim$1--10\% and $\sim$10--100\% for the UFOs and the jet, respectively. Therefore, both of them are able to exert a concurrent feedback impact on the surrounding environment (e.g., Hopkins \& Elvis 2010; Gaspari et al.~2011). 
However, an interesting point to make here is that the powerful and collimated jet tends to ``drill'' out of the galaxy and deliver energy and momentum to large distances. On the other hand, the slower, wider angle and massive UFOs may be much more effective at impacting the structures of the host galaxy (e.g., Tombesi et al.~2012). In fact, from the fraction of sources with detected UFOs Tombesi et al.~(2010a, b) estimated that these absorbers cover a significant fraction of the solid angle, $\sim$0.4. This corresponds to an opening angle of the outflow with respect to the polar axis of $\sim$60$^{\circ}$. Instead, Jorstad et al.~(2005) derived an intrinsic half opening angle of the jet in 3C~111 of only $\sim$3$^{\circ}$. This indicates that the jet covers only a fraction of $\sim$0.001 of the solid angle, which corresponds to less than 1\% of that of the UFOs.

The lower limit of the density of the material in the UFOs can be roughly derived as $n$$=$$N_\mathrm{H} / \Delta r$$\ga$$N_\mathrm{H}/r$. This is valid for compact absorbers ($\Delta r/r$$\la$1) and is supported by the detection of short time-scale variability (e.g., Braito et al.~2007; Cappi et al.~2009; Tombesi et al.~2010a, 2011b). 
For both u7 and u10 we obtain $n$$\ga$$10^7$~cm$^{-3}$. From the photo-ionization code {\sc Xstar} used for the modelling of the Fe~XXV--XXVI absorption lines in Tombesi et al.~(2010b, 2011a, 2011b) we obtain a typical temperature of the plasma of $T$$\sim$$10^6$--$10^7$~K. Therefore, the lower limit on the gas thermal pressure in the UFOs is $P_{\mathrm{th}}$$\simeq$$nkT$$\ga$0.001--0.01~dyne~cm$^{-2}$. This can be even higher considering the significant turbulent velocities of $\sim$1,000~km/s observed for UFOs (Tombesi et al.~2011a, 2011b).
Using the estimate of the jet kinetic power of $\dot{E}_{\mathrm{K, j}}$$\sim$$10^{44}$--$10^{45}$~erg~s$^{-1}$ and assuming a quasi-instantaneous, symmetrical injection of energy, we can derive a crude lower limit of the jet ram pressure at the location of the UFOs of $P_{\mathrm{th, j}}$$\simeq$$\dot{E}_\mathrm{K}/r^3$$\ga$0.001--0.01~dyne~cm$^{-2}$.
These two estimates are comparable, suggesting that accretion disc outflows in the form of UFOs may actually help collimate the inner jet. This also suggests that the initial jet material would encounter more radial than vertical resistance, providing a preferential direction of propagation and this could lead to a ``nozzle-like'' geometrical configuration, which would again help collimate/confine the inner jet (e.g., Blandford \& Rees 1974).

In this regard, we note that recent detailed observations of the inner radio jet of M87, the closest powerful radio-loud AGN to us, revealed that the jet formation is already taking place at distances down to $\sim$10--20$r_\mathrm{s}$ from the super-massive black hole (Hada et al.~2011). Moreover, in the inner few $\sim$100$r_\mathrm{s}$ region the jet is seen opening widely, at an angle of $\sim$$60^{\circ}$, and having a paraboloidal shape. Then, it squeezes down to $\sim$$5^{\circ}$ only at distances of $\sim$$10^5$$r_\mathrm{s}$, after which it transits into a conical shape and becomes a free stream (Asada \& Nakamura 2012). This suggests that the jet is probably subject to an initial collimation by the external gas.
The simultaneous presence at sub-pc scales of these two components in rough pressure equilibrium: a inner, highly relativistic jet, and an outer, more massive, mildly relativistic plasma, is overall consistent with the picture of a transverse stratification of the flow (e.g., Ghisellini et al.~2005; Xie et al.~2012). The line-of-sight inclination angle of 3C~111 is in fact $\sim 18^{\circ}$ (Jorstad et al.~2005). Therefore, this eventuality should be considered when performing numerical simulations of the jet formation.

Theoretically, the complex coupling between radiation, magnetic fields and matter that should be considered to properly explain the formation of outflows/winds from accretion discs has not been accurately solved yet. However, simulations show that disc outflows are ubiquitously produced and can be accelerated to mildly relativistic velocities by radiation and/or magnetic forces (Blandford \& Payne 1982; King \& Pounds 2003; Proga \& Kallman 2004; Ohsuga et al.~2009; Fukumura et al.~2010). 
Moreover, several MHD jet models predict that the jet production is accompanied by the formation of broad, mildly relativistic disc outflows (Blandford \& Payne 1982; Kato et al.~2004; McKinney \& Gammie 2004; McKinney 2006; Tchekhovskoy et al.~2011), potentially providing a direct link between these components.
However, even if MHD models alone could, in principle, already explain the formation of the jet and mildly relativistic disc outflows, the high radiation throughput of AGNs can not be neglected and radiation pressure must play an important role as well (King \& Pounds 2003; Everett \& Ballantyne 2004; Everett 2005; Ohsuga et al.~2009; Ram{\'{\i}}rez \& Tombesi 2012). The comparison with the similar outflows detected in radio-quiet AGNs, which however are known to harbor weak jets as well (e.g., Giroletti \& Panessa 2009; Maitra et al.~2011), may help to clarify this point. This is postponed to a future work.

Comparing the periods of X-ray dips and the observations of UFOs and jet knots in Fig.~1 and Table~2 we note that UFO u7 was observed between the X-ray dip d7 and the jet knot k7. Similarly, u10 was observed between d10 and k10. 
This evidence is intriguing and may suggest the placement of UFOs within the known dip-ejection cycles, which is the most solid observational proof of the disc-jet connection (Marscher et al.~2002; Chatterjee et al.~2009, 2011).
As already discussed by Ch11, the X-ray dip-ejection connection suggests that a decrease in the X-ray production is linked to an increase in the speed of the jet flow, causing a shock front to eventually form and move downstream.
The actual physical cause of this relation is currently still a matter of considerable speculation, however, it has similarities with the stellar-mass black holes (e.g., Livio et al.~2003; Neilsen \& Lee 2009; King et al.~2012), for which more detailed studies have been performed. 
Thermal instabilities in the inner accretion disc can cause a state transition between a radiatively efficient to a low-efficient accretion mode during the X-ray dips, such as in the ADAF/ADIOS regimes (Narayan \& Yi 1995; Blandford \& Begelman 1999), which predicts that a sizeable fraction of the accretion power is not radiated away and instead is released in the form of a jet or outflow. In particular, the jet production mechanism has been demonstrated to be more efficient for both high black hole spins and thick, ADAF-like, inner accretion disks (Meier 2001; Nemmen et al.~2007).  

We note that Tombesi et al.~(2011b) detected an ionized Fe K disc line in the X-ray spectrum of the third observation in Table~1. This observation occurred during a short steady/decreasing flux period after the X-ray dip d10 and no clear UFO was detected. Instead, a blue-shifted Fe XXVI absorption line indicative of a UFO was detected in the successive fourth observation, $\sim$7 days after, during a clear periods of ascending flux. The 2--10~keV S/N of the third observation ($\simeq$105) was enough to detect an absorption line with the same EW as in the fourth (S/N$\simeq$110) if present and its constancy between these two observations was excluded at the 99.9\%. Therefore, the lack of detection of a UFO in the third observation points to the conclusion that it was probably not present along the line-of-sight at that time, but we can not rule out also the possibility that the non-detection was due to strong inhomogeneities/turbulence in the flow. As already stated in Sec. 2.2, we note that the identification of the third observation with a steady/decreasing flux state is less clear because it happened close to a sudden spike in flux and we adopted a conservative approach not considering it in the discussion of the possible relation between the UFOs and period of increasing flux. 
Tombesi et al.~(2011b) interpreted these observations as the evidence of a connection between thermal/structural instabilities in the disc, possibly related to an ADAF state (Wang, Cheng \& Li 2012), and the formation of powerful winds/outflows. 
Interestingly, even if the launching mechanism(s) of the UFOs might not be the same as for the jet, the fact that these observations occurred within two clear dip-ejection cycles suggests that they could be related, in the sense that the strong disc/radiative instabilities that are known to accompany the ejection of a new jet knot could then also trigger or boost the production of disc outflows (Livio et al.~2003; Xie et al.~2012). We note that a similar qualitative behavior, whereby the ejection of jet knots during X-ray dips is followed by an increased production of disc outflows during the successive rising/flaring periods, has been recently reported by Neilsen et al.~(2012) regarding the Galactic microquasar GRS~1915$+$105 during the $\beta$ state.

The main point that we would like to make here is to introduce the notion that UFOs could preferentially appear during phases of increasing flux and possibly at certain times of disc-jet activity. However, given the very limited number of observations, the statistical significance of these relations is only tentative and additional observations are needed to test this properly\footnote{In this regard, we are submitting a large monitoring campaign focused on 3C~390.3, another BLRG with a UFO detection (Tombesi et al.~2010b) and showing the typical jet dip-ejection cycles. The choice fell on this target because it is continuously visible by all main X-ray observatories all year round. The length of the campaign is 20 months and we request a \emph{Swift} monitoring every $\sim$10~days, a long 100ks \emph{Suzaku} and \emph{XMM-Newton} observation every 2 months, a long 100ks \emph{Chandra} HETG observation every 4 months and a parallel VLBA monitoring at intervals of 3 months. If approved, this will allow to have enough observations to conclusively test the relation between UFOs and the jet dip-ejection cycles.}.
Nonetheless, we can certainly conclude that there are now plenty of theoretical and observational evidences that mass and energy loss in the form of winds/outflows from the accretion disc are likely to be the norm rather than the exception and models attempting to explain the link between the jet and the accretion process should take these components into account.

\section{Conclusions}

In this paper we compare the characteristics of the X-ray detected accretion disc outflows, also referred as UFOs, and the radio jet in the broad-line radio galaxy 3C~111. This is the first time that such a study is performed for an AGN. We find that these two outflows, an inner relativistic one and another broader and mildly-relativistic, coexist on sub-pc scales, possibly in agreement with a transverse stratification of a global flow. The two are also potentially in pressure equilibrium, providing additional support for the collimation of the jet. The disc outflows are much more massive than the jet but probably their kinetic power is lower. However, their momentum flux is comparable and both of them are capable to exert a concurrent feedback impact on the surrounding environment. Even though a link between these component is already naturally predicted by MHD jet/outflows simulations, we note that also radiation pressure must be taken into account for a more realistic modelling. The comparison of the detection times of UFOs with the long-term RXTE light curve suggests that they are preferentially observed during periods of increasing flux and the investigation of the VLBA images points to the possibility of placing also these events within the known X-ray dip-ejection cycles, which is the direct evidence of the disc-jet interaction. However, given the limited number of observations, these possible relations are only tentative and additional observations are needed to test them properly. If confirmed, this could provide a new window for X-ray spectroscopy to study processes related to the jet activity on scales even smaller than the $\sim$0.1 milliarcsec reachable with VLBA images.

\section*{Acknowledgments}

FT thank K. Fukumura, R. Nemmen, D. Kazanas, R. F. Mushotzky, A. Tchekhovskoy and S. Dalena for useful discussions. The research at Boston University was funded in part by National Science Foundation grant AST-0907893. The authors thank the anonymous referee for suggestions that led to improvements in the paper.

\end{document}